\documentclass[aps,prc,preprint,groupedaddress]{revtex4}
\usepackage{amsmath,graphicx}

\begin{document}


\title{Elliptic flow from final state interactions in the distorted-wave emission-function model}

\author{Matthew Luzum}
\email{mluzum@phys.washington.edu}
\author{Gerald A. Miller}
\email{miller@phys.washington.edu}

\affiliation{Department of Physics, University of Washington\\
  Seattle, WA 98195-1560}

\date{\today}


\preprint{NT-UW-08-20}

%

%
\begin{abstract}
The effect of final state interactions on the elliptic flow coefficient measured in relativistic heavy ion collisions is investigated within the DWEF formalism established by Miller and Cramer [Phys. Rev. Lett. 94, 102302 (2005),
 J. Phys. G34, 703 (2007), Phys. Rev. C78, 054905 (2008)].  It is found that the optical potential previously found to give the best fit of particle multiplicity and Hanbury Brown and Twiss radii in RHIC events has a moderate effect on the resulting elliptic flow coefficient $v_2$.  This indicates that final state interactions should be taken into account to confidently predict $v_2$ to better than $\sim20$\% accuracy.
\end{abstract}
\maketitle
\section{Introduction}
Hydrodynamic models have been quite successful at describing single particle hadron spectra measured in experiments at the Relativistic Heavy Ion Collider (RHIC) \cite{Teaney:2001av,Huovinen:2001cy,Kolb:2001qz,Hirano:2002ds,Kolb:2002ve}.  Recently, progress has been made in going beyond simple ideal hydrodynamic models and using viscous hydrodynamics to determine the value of various transport coefficients in the RHIC ``fireball,'' e.g.\ the shear viscosity \cite{Luzum:2008cw,Dusling:2007gi,Song:2007ux,Huovinen:2008te}.  Of particular interest is the azimuthal anisotropy of produced particles.  The existence of strong elliptic flow at RHIC is one of the best indications of thermalization at RHIC \cite{Arsene:2004fa,Back:2004je,Adams:2005dq,Adcox:2004mh}, and its precise value is sensitively connected to the viscosity of the medium \cite{Luzum:2008cw}.

Not every aspect of these hydrodynamic models, however, is completely physically justified.  Due to uncertainty in especially early and late time dynamics in the evolution of a heavy ion collision, the initial and final conditions for hydrodynamic evolution are treated somewhat simplistically.  In particular, the so-called Cooper-Frye freezeout algorithm \cite{Cooper:1974mv} is typically implemented to describe how the medium transitions from hydrodynamic behavior to the essentially free particles that enter the detectors.  Using this procedure, the medium, once it reaches some defined freezeout temperature, instantaneously ``freezes out'' from a hydrodynamic fluid to completely free particles that stream to the detectors.  Knowing that this is not entirely physical, it makes sense to investigate how much this treatment might affect the conclusions reached from using these hydrodynamic models.

To that end, Miller, Cramer et al. investigated the effect of introducing some final state interactions to this freezeout behavior in the form of a one-body optical potential that the emitted particles interact with as they ``fight'' to escape the medium \cite{Cramer:2004ih,Miller:2005ji,Luzum:2008tc}.  The main motivation behind this previous work was the notorious inability of hydrodynamic models to fit two-particle correlation data (so-called Hanbury Brown and Twiss radii) while simultaneously fitting single-particle data \cite{Heinz:2002un}. (For a review see Ref.~\cite{Lisa:2005dd}.) 
Thus only multiplicity and HBT radii data were calculated, while the present work extends that to now investigate the effect of these final state interactions on measured elliptic flow.  For simplicity only pions---the dominant hadron produced in a RHIC event---are considered.

The distorted-wave emission-function (DWEF) formalism is briefly reviewed in Sec.~\ref{sec:formalism}, with the calculation of $v_2$ in this formalism described in Sec.~\ref{sec:calculation}.  Sections \ref{sec:results} and \ref{sec:conclusion} contain the results and conclusions.  Details of the calculation are included in Appendix \ref{details} for those interested, along with a semi-analytic derivation of a simple test case used to test the numerics, in Appendix \ref{analytic}.
\section{DWEF Formalism}
\label{sec:formalism}
\label{DWEF}
The DWEF formalism was established in Ref.~\cite{Cramer:2004ih} and described extensively in Ref.~\cite{Miller:2005ji}, and the relevant parts are briefly summarized as follows.  

The main quantity that we are interested in is the detected particle momentum spectra
\begin{equation}
E \frac {dN} {d^3p} = \frac {dN} {dY\ d^2p} = \int d^4x\; S(p,x)
\end{equation}
from which $v_2$ is defined as the second moment in the azimuthal momentum angle
%
\begin{equation}
 v_2 = \langle \cos(2 \phi_p) \rangle =  \frac {\int d\phi_p \; \cos(2 \phi_p) \frac {dN} {dY\ d^2p}.}
{\int d\phi_p \;\frac {dN} {dY\ d^2p}.}
\end{equation}
with $p$ the momentum of the detected particle and $Y$ the particle rapidity.  $\phi_p$ is the angle of the particle momentum with respect to the collision plane.

$S(p,x)$ is known as the emission function.  In conventional hydrodynamical models, it resides in space on a freezeout hypersurface defined by a surface of constant temperature (or other thermodynamic quantity) in the hydrodynamic simulation, with a momentum distribution at each point on the surface given by the appropriate equilibrium (or off-equilibrium in the viscous case) distribution at that given temperature.  This freezeout hypersurface represents the surface of last scattering, from which free particles are emitted and travel directly to the detector.

Here, instead of running a full hydrodynamic simulation, we follow Ref.~\cite{Retiere:2003kf} and use an analytical parametrization of the freezeout surface, similar to one typically found in numerical hydro simulations, but with tunable parameters.  In addition it is allowed to be a more general volume, with finite width in all dimensions, rather than the infinitely thin surface obtained in a conventional Cooper-Frye prescription.

Secondly, instead of freely streaming, particles that are emitted from this surface are then made to interact with an optical potential representing interactions with the medium from which the particles are escaping.

Explicitly we have (see Ref.~\cite{Miller:2005ji} for details)
\begin{equation}
S(p,x) = \frac{\cosh \eta}{(2 \pi)^3}\; e^{\frac{-\eta^2}{2\Delta\eta^2}} \frac{1}{\sqrt{2\pi\; \Delta\tau^2}}\; e^{\frac{-(\tau - \tau_0)^2}{2 \Delta\tau^2}} \frac{M_{\perp}\: \rho (\textbf{b})}{e^{(p \cdot u - \mu_{\pi})/T}-1}
\vert \psi_p^{(-)}(x)\vert^2.
\end{equation}
$p$ is the asymptotic pion momentum, and $M_\perp = \sqrt{\textbf{p}_{\perp}^2 + m_{\pi}^2}$.  As usual, instead of Cartesian coordinates $(t,x,y,z)$, we use the set $(\tau,x,y,\eta)$ or $(\tau,b,\phi,\eta)$ with
\begin{align}\begin{split}
\eta =\ &{\rm arctanh}(z/t) \\
\tau =\ &\sqrt{t^2-z^2} \\
b =\ &\sqrt{x^2 + y^2} \\
\phi =\ &\arctan(y/x).\\
\textbf{b} =\ &(b,\phi).
\end{split}\end{align}
$z$ is the beam direction, with the $xz$ ($\phi=0,\pi$) plane the reaction plane.

$\psi_p^{(-)}(x)$ are the aforementioned distorted waves (as opposed to plane waves appropriate in the absence of interactions).  They obey the equation of motion
\begin{equation}
 \label{waveequation}
 \left( \nabla^2 - \frac{\partial^2}{\partial t^2}-U(\textbf{b})-m_{\pi}^2\right) \psi_p^{(-)}(x) = 0
\end{equation}
for pions interacting with optical potential
\begin{equation}
 U(\textbf{b}) = -(w_0+w_2\ \textbf{p}^2)\ \rho(\textbf{b}).
\end{equation}

Note that although the medium is time-dependent in principle, for simplicity the optical potential is taken here to be time-independent and can be interpreted as a time-averaged quantity.

Whereas in the original DWEF formalism only a rotationally symmetric transverse density $\rho(b)$ was needed (corresponding to central collisions), here we are interested in azimuthal anisotropy and so we need to consider a more general form.  Specifically we take the modified Woods-Saxon profile from Ref.~\cite{Retiere:2003kf}
\begin{equation}
\label{rho}
 \rho (\textbf{b}) = \frac{(\exp[(-1) \frac{R_{ws}}{a_{ws}}] + 1)^2}{(\exp[(b \sqrt{\frac{\cos^2 \phi}{R_x^2} + \frac{\sin^2 \phi}{R_y^2}}-1) \frac{R_{ws}}{a_{ws}}] + 1)^2},
\end{equation}
with $R_{ws} = \sqrt{\frac 1 2 (R_x^2 + R_y^2)}$.  Thus lines of constant density in the transverse plane form ellipses with semimajor to semiminor axis ratio $\frac {R_y}{R_x}$.

Lastly we must specify the fluid velocity $u$, for which we again defer to Ref.~\cite{Retiere:2003kf}.  It is parametrized 
using a transverse fluid rapidity $\eta_t(\textbf{b})$
\begin{equation}
u^{\mu} (x) = (\cosh \eta \cosh \eta_t,\ \sinh \eta_t \cos \phi_b,\ \sinh \eta_t \sin \phi_b,\ \sinh \eta \cosh \eta_t).
\end{equation}
The transverse direction is taken to be perpendicular to lines of constant density.  It can be shown that the angle of such a  fluid velocity, $\phi_b$, obeys \cite{Retiere:2003kf}
\begin{equation}
\phi_b (\phi) = \tan^{-1} (\frac{R_x^2}{R_y^2} \tan \phi).
\end{equation}

The transverse fluid rapidity $\eta_t(\textbf{b})$ is first taken to have the same elliptic symmetry as the density, increasing linearly with the ``radial'' coordinate $\tilde{b} \equiv \sqrt{\frac{(b\cos(\phi))^2}{R_x^2}+\frac{(b\sin\phi))^2}{R_y^2}}$.  Then added to this is a term proportional to $\cos(2 \phi)$ representing the amount of elliptic flow built up before freezeout
\begin{equation}
\label{rapidity}
 \eta_t(\textbf{b}) = \eta_f\; b \sqrt{\frac{\cos^2 \phi}{R_x^2} + \frac{\sin^2 \phi}{R_y^2}} (1 + a_2\; \cos(2 \phi)).
\end{equation}

The momentum in these coordinates takes the form
\begin{equation}
p^\mu = (M_\perp \cosh Y , p_\perp \cos \phi_p , p_\perp \sin \phi_p , M_\perp \sinh Y).
\end{equation}
We choose to focus on data at midrapidity, $Y = 0$, and so 
\begin{equation}
 p \cdot u = M_{\perp} \cosh \eta\; \cosh \eta_t - p_{\perp} \sinh \eta_t\; \cos(\phi_b - \phi_p).
\end{equation}

In all, then, the parameters involved in this model are: $\Delta\eta$, $\Delta\tau$, $\tau_0$, $\mu_\pi$, $T$, $w_0$, $w_2$, $R_x$, $R_y$, $a_{ws}$, $\eta_f$, and $a_2$.  We are interested in the effect of an optical potential like the one found to give the best fit in Ref.~\cite{Miller:2005ji} and so we will keep all of these parameters fixed to those best-fit values, and only adjust $\frac {R_y} {R_x}$ and $a_2$ to give reasonable results for non-central collisions.

It should be noted that the formalism developed is not strictly correct when the optical potential is complex.  (See the discussion in Ref.~\cite{Luzum:2008tc}.)  We therefore also investigate the best fit values of Ref.~\cite{Luzum:2008tc} for a vanishing imaginary part of the optical potential.
\section{Calculating $v_2$}
\label{sec:calculation}
This section outlines how the calculations are carried out.  A set of coupled differential equations must be solved numerically to obtain the wavefunctions $\psi_p^{(-)}$, and then a five-dimensional integral must be performed (two of which can be done analytically with suitable approximations.)
\subsection{The Wavefunctions $\psi_p^{(-)}(x)$}
$\psi_p^{(-)}$ satisfies Eq.~(\ref{waveequation}).
Since $U(\textbf{b})$ is independent of $t$ and $z$, we can write
\begin{equation}
 \psi_p^{(-)}(x) \equiv e^{- i \omega_p t} e^{i p_z z} \psi_p^{(-)}(\textbf{b}),
\end{equation}
and Eq.~(\ref{waveequation}) becomes
\begin{equation}
  \left( \nabla_{\perp}^2 - U(\textbf{b})+p_{\perp}^2 \right) \psi_p^{(-)}(\textbf{b}) = 0,
\end{equation}
or
\begin{equation}
  \left( \frac{\partial^2}{\partial b^2} + \frac{1}{b} \frac{\partial}{\partial b} + \frac{1}{b^2} \frac{\partial^2}{\partial \phi^2} - U(\textbf{b})+p_{\perp}^2 \right) \psi_p^{(-)}(\textbf{b}) = 0.
\end{equation}

Decomposing $\psi_p^{(-)}$ and $U(\textbf{b})$ into angular moments
\begin{align}
 \psi_p^{(-)}(\textbf{b}) &= \sum_{m = -\infty}^{\infty} f_m(p,b){(-i)}^m e^{i m\; (\phi - \phi_p)}, \\
\label{Um}
  U(\textbf{b}) &\equiv \sum_n U_n(b) e^{i n \phi},
\end{align}
results in
\begin{equation}
\sum_m \left[  \left( \frac{\partial^2}{\partial b^2} + \frac{1}{b} \frac{\partial}{\partial b} - \frac{m^2}{b^2} + p_{\perp}^2 \right) f_m - \sum_n U_n\ f_{m-n}\; i^n e^{i n \phi_p} \right] e^{i m\phi}e^{-i m \phi_p} = 0.
\end{equation}

So the term in brackets vanishes identically for each $m$, and we must solve a set of coupled differential equations. In practice, every $f_m$ above a certain $m_{max}$ is set to zero, and a finite set of coupled equations is solved numerically.

The boundary conditions are the same as for the cylindrically symmetric case---far outside the medium one should have a canonically normalized plane wave plus an outgoing wave, i.e.
\begin{equation}
\label{boundary}
 f_m (b \gg R_{ws}) = J_m (p\ b) + T_m H_m^{(1)} (p\ b)
\end{equation}
with $J_m$ and $H_m^{(1)}$ Bessel functions and Hankel functions of the first kind, respectively.

Details of this calculation can be found in Appendix \ref{details}.  The program used to calculate the wavefunctions was tested in part by comparing to a semi-analytic solution described in Appendix \ref{analytic}.

\subsection{Integration}

Once the wavefunctions are found, a five-dimensional integral must be performed:
\begin{equation}
\label{v2}
 v_2 =  \frac{\int d\phi_p \; \cos(2\; \phi_p)  \int d^4x\; S(p,x)}{\int d\phi_p  \int d^4x\; S(p,x)}.
\end{equation}
The $\tau$ integral can be done analytically
\begin{equation}
\int \tau d\tau e^{\frac{-(\tau - \tau_0)^2}{2 \Delta\tau^2}} = \sqrt{2 \pi} \tau_0 \Delta\tau.
\end{equation}
The $\eta$ integral can also be done analytically with the following approximations (as in Ref.~\cite{Miller:2005ji})
\begin{align}
e^{\frac{-\eta^2}{2\Delta\eta^2}} \approx&\ e^{\frac{1}{\Delta\eta^2}}e^{-\frac{\cosh \eta}{\Delta\eta^2}} \\
\frac{1}{e^{(p \cdot u - \mu_{\pi})/T}-1} \approx& \sum_{j = 1}^{j_{max}} e^{(-p \cdot u + \mu_{\pi})j/T},
\end{align}
where the Bose-Einstein distribution is approximated by a sum over Boltzmann distributions truncated at some $j_{max}$, and so
\begin{equation}
 \int d\eta\; \cosh \eta\; e^{-\cosh \eta (\frac{1}{\Delta\eta^2}+ \frac{M_{\perp}j}{T} \cosh \eta_t)} = 2 K_1 \left(\frac{1}{\Delta\eta^2}+ \frac{j}{T}M_{\perp}\; \cosh \eta_t \right).
\end{equation}

Finally, then, for the numerator we have
\begin{eqnarray}
\label{integrals}
\lefteqn{\int d\phi_p\;\cos(2 \phi_p) \int d^4x\; S(p,x)} \nonumber \\
 &=&  \frac{2\; \tau_0 M_\perp }{(2\pi)^3}\; e^{\frac{1}{\Delta\eta^2}}\sum_{m,n,j} i^{n-m}e^{\frac{\mu_\pi j}{T}} \nonumber \\
&& \times  \int d^2b\; \rho(\textbf{b}) f_m(p,b)\;f_n^*(p,b)\; e^{i(m-n)\phi} K_1 \left(\frac{1}{\Delta\eta^2}+ \frac{j}{T}M_{\perp}\; \cosh \eta_t \right)\nonumber \\
&& \times  \int d\phi_p\;\cos(2 \phi_p) e^{-i(m-n)\phi_p} e^{\frac j T p_\perp \sinh(\eta_t)\cos(\phi_b-\phi_p)},
\end{eqnarray}
and similarly for the denominator.  The final three integrals are done numerically.

More details of this part of the calculation can also be found in Appendix \ref{details}.
\section{Results}
\label{sec:results}
\begin{table*}
  \caption{Best fit parameter sets.  The top line (Fit 1) is a general fit \cite{Miller:2005ji} while the bottom line (Fit 2) is from a fit where $Im(w_2)$ is held at 0.0001 \cite{Luzum:2008tc}. \label{table}}
  \vspace{0.1cm}
  \begin{tabular}{|c|cccccccccc|}\hline
  & $T$ & $\eta_f$ & $\Delta\tau$ &$R_{WS}$ & $a_{WS}$& $w_0$& $w_2$ &$\tau_{0}$ &$\Delta\eta$ & $\mu_\pi$\\
  & $(MeV)$ & & $(fm/c)$ &$(fm)$ & $(fm)$& $(fm^{-2})$& & $(fm/c)$ & & $(MeV)$\\
\hline
 Fit 1: & 156.58 & 1.310 & 2.0731 & 11.867 & 1.277 & 0.0693 & 0.856+$i$0.116 & 9.04 & 1.047 & 139.57\\
 Fit 2: & 121 & 1.05 & 0 & 11.7 & 1.11 & 0.495 & 0.762+$i$0.0001 & 9.20 & 70.7 & 139.57\\
     \hline
 \end{tabular}
\end{table*}
As previously mentioned, we would like to determine the effect of adding final state interactions to hydrodynamic fits.  To gain insight into this, we consider an emission function with parameter values taken from Refs.~\cite{Miller:2005ji,Luzum:2008tc}, which give the best description of the single particle data in general, and also with the imaginary part of the optical potential held at zero (see Table~\ref{table}.  Also note that in both fits the chemical potential was fixed at the pion mass).  

We must make alterations to this central collision model to approximate a more peripheral collision.  The results for a central collision do not unambiguously imply what a peripheral collision will look like without appealing to a particular model for the dynamics of the system.  We therefore choose reasonable parameters to approximately represent a collision with impact parameter $\sim 7$ fm, and then see how the resulting $v_2$ depends on the strength of the optical potential.  In principle one could vary all the parameters and do a separate fit of all the relevant experimental data (multiplicity, HBT radii, $v_2$, etc.) for each of various collision centralities.  However, the computing time to do so would be prohibitive, and here we are most interested in investigating only the effect of the interactions, so we proceed as follows.

First, as in Ref.~\cite{Miller:2005ji}, we scale down $R_{ws}$, $a_{ws}$, and $\tau_0$ by the number of participants to the 1/3 power, with $N_{part}$ taken from the Glauber model (with the same parameters used in Ref.~\cite{Luzum:2008cw}) for an impact parameter of 0 and 7 fm ($N_{part} = 377.5$ and $171.544$).  Specifically $R_{ws} \rightarrow 0.7688 R_{ws}$.   Then we adjust the ratio $\frac{R_y}{R_x}$ such that the spatial eccentricity
\begin{equation}
\label{ecc}
 \epsilon \equiv \frac{\langle y^2\rangle - \langle x^2\rangle}{\langle y^2\rangle + \langle x^2\rangle} = \frac{R_y^2 - R_x^2}{R_y^2 + R_x^2}
\end{equation}
has a value of 0.035.  This is a reasonable value corresponding to the spatial eccentricity at freezeout of hydrodynamic fits of peripheral collisions with impact parameter $\sim7$ fm.  Note that the brackets in Eq.~\ref{ecc} indicate a spatial average with weight given by Eq.~\ref{rho}, while the spatial eccentricity in hydrodynamic simulations are typically given with respect to , e.g., energy density.  We nevertheless keep the eccentricity from Eq.~\ref{ecc} fixed at this value with an understanding that it is only a rough but still realistic guide to the shape.

\begin{figure}
 \includegraphics{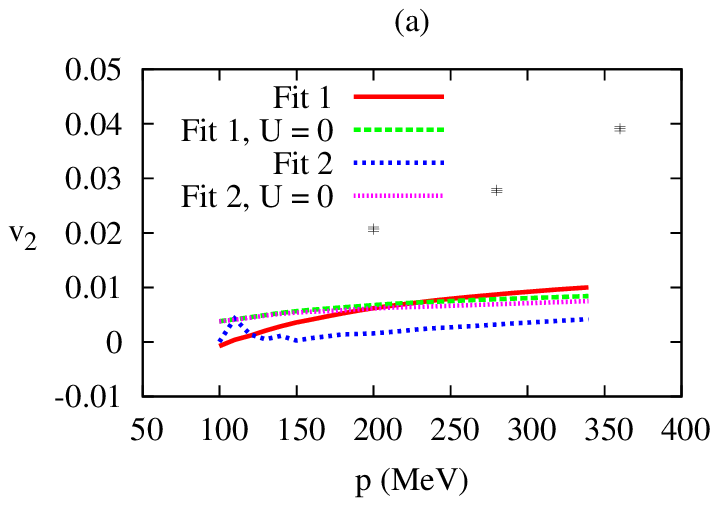}
 \includegraphics{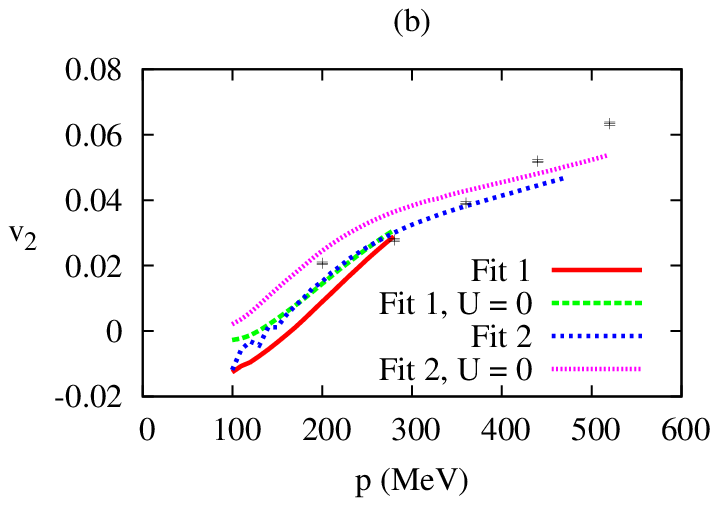}
 \caption{\label{a0}(Color online) Calculated $v_2$ as a function of momentum with $a_2 = 0$ (a) and $a_2 = 0.11, 0.10$ (Fit 1, 2) (b).  Points with error bars are experimental data for pions at 20--30\% centrality from the STAR Collaboration \cite{Adams:2004bi}.}
\end{figure}

Lastly we must specify how much elliptic fluid flow is built up in earlier stages of the collision, represented by the value of $a_2$ (recall Eq.~\ref{rapidity}).  First we set $a_2 = 0$ and see what $v_2$ is generated by interactions with the optical potential in the absence of significant elliptic fluid flow (Fig.~\ref{a0}(a)).  The calculated elliptic flow coefficient $v_2$ is plotted as a function of momentum, along with the relevant experimental data.  (Note that $\textbf{p}$ in our calculation is the momentum of an asymptotically free pion detected far outside the medium, not the momentum of a particle as it is emitted inside the medium, and can therefore be compared directly to experiment.)  Although we are only able to calculate up to a limited momentum, it is clear that final state interactions alone do not generate an appreciable value for $v_2$ for either the general best-fit parameters (Fit 1) or those with a vanishing imaginary part of the optical potential (Fit 2).



Next we increase $a_2$ such that the experimental value for $v_2$ is roughly obtained  (Fig.~\ref{a0}(b)).  A value of $a_2 = 0.11$ was required for the parameters from Fit 1, while $a_2 = 0.10$ was sufficient to bring the emission function from Fit 2 into the physical regime.  One can see that the optical potential has a small but non-negligible effect---it decreases $v_2$ on the order of 10--25\% of its zero-interaction value with a slightly smaller effect as momentum increases.
\section{Conclusion}
\label{sec:conclusion}
Final state interactions in the DWEF model were found to have a small, though not entirely insignificant effect on the elliptic flow coefficient $v_2$.  
This is in addition to the indirect effect of adding final state interactions.  For example, adding an optical potential changes other observables such as the multiplicity, which would alter parameters in a hydrodynamic fit such as freezeout temperature, which would then in turn have an effect on the calculated value of $v_2$.

The precise size of these effects in general can only be determined with a better understanding of the model fits (e.g. Fit 1 versus Fit 2) in addition to a more detailed analysis---a full parameter search using all the relevant experimental data, or perhaps even by adding final state interactions directly into hydrodynamic simulations
(i.e. a hydrodynamic afterburner in the vein of, e.g., Refs.~\cite{Bass:2000ib, Hirano:2005xf, Nonaka:2006yn, Teaney:2001av, Petersen:2008dd}).  It is reasonable, however, to conclude that final state interactions can affect the calculated value of $v_2$ by as much as $\sim20$\% (in agreement with other investigations of final state interactions, e.g., Ref.~\cite{Teaney:2001av}), and so must be properly taken into account to have confidence in the quantitative predictions of hydrodynamic simulations at that level of precision.
%
%
%
%
\begin{acknowledgments}
We thank Paul Romatschke and John Cramer for helpful discussions.  This work was supported by the US Department of Energy, grant 
numbers DE-FG02-97ER41014 and DE-FG02-00ER41132.
\end{acknowledgments}

\appendix

\section{Details of the Numerical Implementation}
\label{details}
A program was written in C++, making use of the GNU Scientific Library (GSL) version 1.9, to do the calculation of $v_2$, as detailed here.

The integral over the azimuthal angle of the pion momentum, $\phi_p$ is done as a sum using a simple trapezoid rule.  This is because for each different value of $\phi_p$, a new set of differential equations must be solved.  This also allows for the numerator and denominator of Eq.~\ref{v2} to be solved simultaneously, with just a factor of $\cos(2 \phi_p)$ multiplied to the numerator when adding terms to the sum.

For each term in the sum, then, first the wavefunctions $\psi_p^{(-)}$ are obtained.  They obey a set of coupled differential equations of the form
\begin{equation}
\left( \frac{\partial^2}{\partial b^2} + \frac{1}{b} \frac{\partial}{\partial b} - \frac{m^2}{b^2} + p_{\perp}^2 \right) f_m - \sum_n U_n\ f_{m-n}\; i^n e^{i n \phi_p} = 0
\end{equation}
for all integers $m$.  This set is truncated, since large $m$ moments ($f_m$ for $m > p_\perp R_{ws}$) contribute little to the wavefunction.  Therefore, all $f_m$ for $m$ greater than some $m_{max}$ are set to zero, leaving a finite ($2 m_{max} + 1$) number of coupled ordinary differential equations.  These are solved by calling a GSL solver.  Using an embedded Runge-Kutta-Fehlberg method seemed to give the best performance.  For these solutions, Eq.~\ref{rho} is integrated numerically to find the moments $U_n$.  This is done with the GSL adaptive integration routine for oscillatory functions.

To match to the proper boundary conditions, one must find ($2 m_{max} + 1$) linearly independent solutions to this set of equations and take the correct linear combination of these solutions that matches the desired boundary conditions.  The straightforward choice for these linearly independent solutions is to sequentially solve for the case where only one of the partial waves is non-zero near the origin.  For example, for the n'th solution let:
\begin{eqnarray}
 f_m (b = b_{min} << \frac 1 p) &=& \delta_{m,n} \nonumber \\
 f_m'(b_{min}) &=& \frac m b \delta_{m,n}
\end{eqnarray}
and then solve the set of differential equations up to some arbitrarily large $b_{max}$ far outside the potential. We can then match each partial wave in this $n^{th}$ solution to the form:
\begin{equation}
 f_{m,n} (b_{max}) =  A_{m,n} J_m (p\ b) + B_{m,n} H_m^{(1)} (p\ b).
\end{equation}

The final wavefunction is then given by the linear combination of these solutions that matches the form of Eq.~\ref{boundary} at $b_{max}$:
\begin{equation}
 f_m (b) = \sum_n C_n f_{m,n}(b).
\end{equation}

This part of the program was tested with the trivial case of zero optical potential, in addition to comparing to a separately written program that calculates only the cylindrically symmetric case, as well as to the results of the semi-analytical test case described in Appendix \ref{analytic}.
 
Once these wavefunctions are obtained and stored in memory, the integral over $b$ and $\phi$ in Eq.~\ref{integrals} can be performed in addition to the sum over Boltzmann factors.  The integrations are done with two GSL adaptive integration routines, one embedded in the other.  The sum is done inside the argument of the integrals.
\section{Semi-Analytic Test Case}
\label{analytic}
To test the numerics, the case of a pion moving through an elliptically-shaped step-function potential was solved (semi-)analytically making use of elliptic coordinates.  This can be compared to the case of $a_{ws}\rightarrow 0$ (see Sec.~\ref{DWEF}).

We want to solve Eq.~\ref{waveequation} with $U{(\textbf{b})}$ an elliptically shaped step function---a finite potential inside an ellipse in the transverse plane, with zero potential outside.

It is useful to change to elliptic (cylindrical) coordinates, denoted $u$ and $v$.  Think of $u$ as a 'radial' coordinate that runs from 0 to $\infty$ and $v$ as an 'angular' coordinate that runs from 0 to $2 \pi$
\begin{eqnarray}
x = a\ \cosh(u)\ \cos(v) \nonumber \\ 
y = a\ \sinh(u)\ \sin(v).
\end{eqnarray}
Note the major and minor axes of the resulting confocal ellipses are reversed from the shape of the density used in the main calculation (which is larger in the $y$ direction).  This is to maintain consistency with the conventional definition of elliptic coordinates.   At the end one can simply take $\phi_p \to (\phi_p + \pi)$ to match the usual convention in RHIC papers.

Consider the case
\begin{equation}
U(\textbf b) = U(u) = U_0\ \Theta(u_0 - u).
\end{equation}
The sharp boundary at $u = u_0$ is an ellipse with major and minor axes
\begin{eqnarray}
R_x = a\ \cosh(u_0) \nonumber \\
R_y = a\ \sinh(u_0).
\end{eqnarray}

In this coordinate system the Laplacian is
\begin{equation}
\nabla^2_\perp = \frac{1}{a^2 \left( \sinh^2(u)+\sin^2(v) \right)} \left( \frac{\partial^2}{\partial u^2} + \frac{\partial^2}{\partial v^2} \right)
\end{equation}
and so Eq.~(\ref{waveequation}) becomes
\begin{equation}
\left[ \frac{1}{a^2 \left( \sinh^2(u)+\sin^2(v) \right)} \left( \frac{\partial^2}{\partial u^2} + \frac{\partial^2}{\partial v^2} \right) - U(u) + p^2 \right] \psi_p ({\bf b}) = 0
\end{equation}
or equivalently
\begin{equation}
\left[ \frac{\partial^2}{\partial u^2} + 2q(u)\ \cosh(2u)  + \frac{\partial^2}{\partial v^2} - 2 q(u)\ \cos(2v) \right] \psi_p ({\bf b}) = 0
\label{el}
\end{equation}
with 
\begin{equation}
q(u) = \frac{a^2}{4} \left( p^2 - U(u) \right).
\end{equation}
On the inside of the potential and on the outside separately, $q(u)$ does not depend on $u$ and these cases can be solved with separation of variables and the solutions patched together at $u = u_0$.
Let
\begin{eqnarray}
q_{in}& = &\frac{a^2}{4} \left( p^2 - U_0 \right) \nonumber \\
q_{out}& = &\frac{a^2}{4} p^2.
\end{eqnarray}

Start by expanding $\psi_p ({\bf b})$ in terms of so-called elliptic sines and cosines of the 'angular' variable $v$.  They are solutions of `Mathieu's equation' \cite{gradshteyn_table_2000}:
\begin{equation}
\left( - \frac{\partial^2}{\partial v^2} + 2 q\ \cos(2v) \right) C(\alpha,q,v) = \alpha \ C(\alpha,q,v) \label{Mathieu}.
\end{equation}
The general solutions are called `Mathieu functions,' usually denoted $C(\alpha,q,v)$ for solutions even in the coordinate $v$ and $S(\alpha,q,v)$ for odd.  Demanding periodicity of the variable $v$ allows only certain discreet eigenvalues $\alpha$ (denoted here $\alpha_n$ for the even functions and  $\beta_n$ for the odd functions).  This (complete) set of periodic solutions is commonly called elliptic sines and elliptic cosines:

\begin{eqnarray}
C(\alpha_n,q,v) \equiv ce_n (v,q) \nonumber \\
S(\beta_n,q,v) \equiv se_n (v,q).
\end{eqnarray}

The general solution of Eq.~(\ref{el}) can be written in terms of these elliptic sines and cosines:
\begin{equation}
\psi_p ({\bf b}) \equiv \sum_{n=0}^\infty \left[ f_{c_n}(u) ce_n (v,q) + f_{s_n}(u) se_n (v,q) \right].
\end{equation}

Plugging this in to Eq.~(\ref{el}) gives
\begin{eqnarray}
\left[ \frac{\partial^2}{\partial u^2} + 2q\ \cosh(2u)  - \alpha_n \right] f_{c_n} (u) = 0 \\
\left[ \frac{\partial^2}{\partial u^2} + 2q\ \cosh(2u)  - \beta_n \right] f_{s_n} (u) = 0.
\end{eqnarray}

This is called the modified Mathieu equation, which can be obtained from Eq.~(\ref{Mathieu}) by replacing $v \to (i\ u)$. Note that the eigenvalues are different for the functions corresponding to $ce_n$ and $se_n$ ($f_{c_n}$ and $f_{s_n}$ above, respectively). The general solution is then the same as for the original Mathieu equation, analytically continued with $v \to (i\ u)$, though typically they are organized by boundary conditions analogous to Bessel and Neumann functions (denoted $Je_n(u,q)$, $Ne_n(u,q)$, etc.) \cite{visual}:

\begin{eqnarray}
f_{c_n} (u) = C_{c_n} Je_n(u,q) + S_{c_n} Ne_n(u,q) \\
f_{s_n} (u) = C_{s_n} Jo_n(u,q) + S_{s_n} No_n(u,q).
\end{eqnarray}

Note that there are many different sets of so-called Mathieu functions, each being a complete orthogonal basis.  Replacing $q_{in}$ with $q_{out}$ results in a different basis, and there are separate sets of modified Mathieu functions corresponding to the eigenvalues of the elliptic sines and elliptic cosines ($\alpha_n$ and $\beta_n$).

By requiring continuity at the $u=0$ line segment one finds that the general solution inside the potential is:
\begin{equation}
\psi_p^{in}(u,v) = \sum_n \left[ Ce^{in}_n Je_n(u,q_{in}) ce_n(v,q_{in}) + Co^{in}_n Jo_n(u,q_{in})se_n(v,q_{in}) \right]
\end{equation}
with undetermined coefficients $Ce^{in}_n, Co^{in}_n$.

Outside, we write the solution as the sum of a plane wave and an outgoing wave \cite{McLachlan:1947}
\begin{eqnarray}
\psi_p^{out}(u,v) = \nonumber \\
\sum_n [ \left( \frac{1}{p_n} Je_n(u,q_{out}) + Ce^{out}_n He_n^{(1)}(u,q_{out})\right) ce_n (v,q_{out}) ce_n(\phi_p,q_{out}) \nonumber \\
+\left( \frac{1}{s_n} Jo_n(u,q_{out}) + Co^{out}_n Ho_n^{(1)}(u,q_{out})\right) se_n (v,q_{out}) se_n(\phi_p,q_{out})],
\end{eqnarray}
where the H's are analogous to Hankel functions
\begin{eqnarray}
He^{(1)}_n (u,q) \equiv Je_n (u,q) + i\ Ne_n (u,q) \\
Ho^{(1)}_n (u,q) \equiv Jo_n (u,q) + i\ No_n (u,q)
\end{eqnarray}
and the plane wave coefficients $p_n$ and $s_n$ are
\begin{eqnarray}
 \frac{1}{p_n}  =  \frac 1 \pi \int_0^{2\pi} dv\ e^{i p\cdot x} ce_n (v,q_{out}) \\
 \frac{1}{s_n}  =  \frac 1 \pi \int_0^{2\pi} dv\ e^{i p\cdot x} se_n (v,q_{out}).
\end{eqnarray}
The coefficients $Ce^{out}_n$ and $Co^{out}_n$, along with the analogous 'inside' coefficients are determined by matching boundary conditions.

To match at the $u = u_0$ boundary, project the 'inside' angular functions\\ (e.g. $ce_n(v,q_{in})$) in terms of the 'outside' ones (e.g. $ce_n(v,q_{out})$).
\begin{eqnarray}
ce_j(v,q_{in}) = \sum_{n=0}^\infty B^c_{jn} ce_n(v,q_{out}) \\
se_j(v,q_{in}) = \sum_{n=0}^\infty B^s_{jn} se_n(v,q_{out}),
\end{eqnarray}
with
\begin{eqnarray}
B^c_{jn} = \frac{1}{\pi} \int_0^{2\pi} dv ce_j(v,q_{in})\ ce_n(v,q_{out}) \\
B^s_{jn} = \frac{1}{\pi} \int_0^{2\pi} dv se_j(v,q_{in})\ se_n(v,q_{out}).
\end{eqnarray}

Then the 'inside' wave functions are
\begin{equation}
\psi_p^{in} = \sum_{j,n} \left[ Ce^{in}_j\ Je_j(u,q_{in})\ B^c_{jn} ce_n(v,q_{out}) + Co^{in}_j\ Jo_j(u,q_{in})\ B^s_{jn} se_n(v,q_{out})\right].
\end{equation}
The coefficients ($Ce^{in}_n$, $Co^{in}_n$, $Ce^{out}_n$, $Co^{out}_n$) can then be determined by demanding that $\psi$ and its gradient be continuous at $u = u_0$, which gives the following relations:

\begin{widetext}

\begin{equation}
\sum_j Ce^{in}_j Je_j(u_0,q_{in}) B^c_{jn} = \frac{1}{p_n} Je_n(u_0,q_{out}) ce_n(\phi_p,q_{out}) + Ce^{out}_n He^{(1)} (u_0,q_{out}) ce_n(\phi_p,q_{out})
\end{equation}
\begin{equation}
\sum_j Co^{in}_j Jo_j(u_0,q_{in}) B^s_{jn} = \frac{1}{s_n} Jo_n(u_0,q_{out}) se_n(\phi_p,q_{out}) + Co^{out}_n Ho^{(1)} (u_0,q_{out}) se_n(\phi_p,q_{out})
\end{equation}
\begin{equation}
\sum_j Ce^{in}_j Je'_j(u_0,q_{in}) B^c_{jn} = \frac{1}{p_n} Je'_n(u_0,q_{out}) ce_n(\phi_p,q_{out}) + Ce^{out}_n He'^{(1)} (u_0,q_{out}) ce_n(\phi_p,q_{out})
\end{equation}
\begin{equation}
\sum_j Co^{in}_j Jo'_j(u_0,q_{in}) B^s_{jn} = \frac{1}{s_n} Jo'_n(u_0,q_{out}) se_n(\phi_p,q_{out}) + Co^{out}_n Ho'^{(1)} (u_0,q_{out}) se_n(\phi_p,q_{out}).
\end{equation}

\end{widetext}

The plane wave coefficients ($p_n,s_n$) as well as the coefficients from the projection ($B^c_{jn}, B^s_{jn}$) must be solved numerically.  In addition, to compare to the $f_m$ in the main calculation, the resulting wavefunctions are integrated to project out the usual angular moments.  Hence the description as a ``semi-analytical'' test case.  In fact, this implementation (done in Mathematica) saves no time over the original numerical version, but it does provide an independent check.

 \bibliography{RHIC}

\end{document}